\documentclass[aps,prl,twocolumn,groupedaddress]{revtex4}

\usepackage{graphicx}
\usepackage{amssymb}

\usepackage{hyperref}

\begin{document}


\title{Isospin conserving Dark Matter with isospin dependent interaction, and reconciliation of contrasting results from direct Dark Matter experiments}


\author{F. Giuliani}
\email[]{fgiulian@unm.edu}
\affiliation{Department of Physics and Astronomy, University of New Mexico, NM, USA}


\date{\today}

\begin{abstract}
A simple model of Dark Matter (DM) which couples with the two nucleons with different coupling strengths without violating isospin conservation is explored as an example of the importance of keeping the experimental data analysis as model independent as possible. The apparent contrast between the recent CRESST event excess and current XENON100 spin-independent (SI) exclusions is shown to be removed by simply not assuming that the couplings of the DM with the two nucleons are equal.
\end{abstract}


\maketitle


The identification of Dark Matter is a long standing problem of modern astroparticle physics and cosmology. A particle candidate to be a constituent of DM needs to be stable on the time scale of the Universe age, have a suitably small annihilation rate at its typical densities, and the strength of its interaction with photons and baryonic matter must be weak enough to explain the absence of clear traces of its presence other than the gravitational effects on astronomical or cosmological scales, like the marked separation of the gravitational lensing from the x-ray and light emission in cluster collisions (\cite{bullcluster}). Moreover, since a high number density would make the annihilation rates high, the DM particle(s) should be massive, reasonably heavier than the proton. Since no particle in the Standard Model of Particle Physics combines all these characteristics, the particle theory of DM has to be an extension of the Standard Model. Experimentally, the search is carried on either indirectly by looking for secondaries of the DM annihilations (like $\gamma$ or positron excesses and $\nu$s), or directly, by seeking rare nuclear scattering events in underground detectors. Heuristically, the lowest order effective scattering interaction has a Spin-Independent (SI) component given, in the nonrelativistic limit, by the product of the number densities, and a Spin-Dependent (SD) component arising, if the DM particle has spin, from the dot product of the spins of the DM particles and the target nuclei.

But, while for the SD component it is generally recognized that the coupling strengths with the two nucleons may be different, for the SI component it has become customary to assume, in the analysis of direct DM experimental results, that the couplings $f_{p,n}$ with the two nucleons are equal. This assumption is a priori arbitrary, restricting the analysis to some models, like the MSSM lightest neutralino with squarks much heavier than the higgs (if the latter condition is not satisfied, the MSSM neutralino interacts with $f_{p}\neq f_{n}$ \cite{mssmrep}). In fact, a general lowest order effective lagrangian for the SI scattering of a DM fermion $\chi$ off nucleons is \cite{kurikamio,me}:

\begin{equation}
{\cal L}\propto \overline{\chi}\chi(f_ {p}\overline{p}p+f_ {n}\overline{n}n)
\end{equation}

\noindent whose scalar operators $\overline{\chi}\chi$, $\overline{p}p$, and $\overline{n}n$ act, in the non-relativistic limit, as particle number density operators. The resulting spin-independent zero-momentum transfer cross section for the elastic scattering of a DM particle off a nucleus is then determined by the particle numbers:

\begin{equation}
\label{sisig}
\sigma\propto \mu^{2}(f_{p}Z+f_{n}(A-N))^{2}
\end{equation}

The experimental reason for reporting limits extracted from a null result assuming $f_{p}=f_{n}$ is that this scenario is the best constrained, or, equivalently, the most easily detectable, by any direct DM experiment, since  the scattering cross section of Eq. (\ref{sisig}) is maximal for any target nucleus. This allows to compare the discovery potential of experiments which only reported limits. But, when comparing limits to a positive signal (or two positive signals), which may arise from a more generic scenario, $f_{p}$ and $f_{n}$ should be left independent of each other. This is known to reduce or even remove the apparent incompatibility of results like CoGeNT and XENON \cite{cogexe, ivdm}, provided that suitable values of $f_{p,n}$ are employed. Moreover, it is possible to constrain \cite{me}, and even determine, $f_{p}$ and $f_{n}$ by combining different experiments. 

A simple example of a DM interaction producing different scattering cross sections on the two nucleons is the following minimal extension of the standard model, that exhibits $f_{p}\neq f_{n}$ while manifestly conserving isospin.  The new (dark) particle content consists of an isospin spinor doublet $\Psi=(\psi,\chi)$, along with a scalar isospin triplet $\vec{\phi}$ and a scalar singlet $\rho$ mediating the dark interaction. The lagrangian is:

\begin{equation}
\label{lagra}
{\cal L}= g_{t}\vec{\phi} \cdot(\overline{\Psi}\vec{\sigma}\Psi +\overline{N}\vec{\sigma}N)+g_{s}\rho(\overline{\Psi} \Psi+\overline{N}N)
\end{equation}

\noindent where $\vec{\sigma}$ are the Pauli matrices for the isospin vectors, and $g_{t,s}$ are the triplet and singlet coupling constants. $N=(p,n)$ is the nucleon doublet. Such a scenario could arise from a weakened strong interaction in the dark sector, resulting in highly massive dark mesons coupled to both the nucleons and the dark fermions with the weakened $g_{s,t}$ constants. If $M_{s,t}$ are the respective masses of the scalar singlet and triplet, the low energy effective lagrangian for elastic scattering is:

\begin{equation}
\label{efflagra}
{\cal L}=\left(\frac{g_{t}}{M_{t}}\right)^{2}\overline{\Psi} \vec{\sigma}\Psi\cdot \overline{N}\vec{\sigma}N +\left(\frac{g_{s}}{M_{s}}\right)^{2} \overline{\Psi} \Psi\overline{N}N
\end{equation}

If we now assume that the dark doublet mass splitting $M_{\psi}-M_{\chi}$ is large relative to the DM kinetic energies, with the lightest dark spinor being $\chi$ (e.g. $M_{\psi}-M_{\chi}\gg 1-10$ MeV, for $1\leq M_{\chi}\leq 1000$ GeV), then the inelastic scattering is suppressed, reducing Eq (\ref{efflagra}) to:

\begin{equation}
\label{efflagra2}
\begin{array}{l}
{\cal L}= -\left(\frac{g_{t}}{M_{t}}\right)^{2}\overline{\chi}\chi \overline{N} \sigma^{3}N + \left(\frac{g_{s}}{M_{s}}\right)^{2} \overline{\chi} \chi(\overline{n}n+\overline{p}p)= \\
-\left(\frac{g_{t}}{M_{t}}\right)^{2}\overline{\chi}\chi (\overline{p}p-\overline{n}n) + \left(\frac{g_{s}}{M_{s}}\right)^{2} \overline{\chi} \chi(\overline{n}n+\overline{p}p)=\\
\overline{\chi} \chi((\frac{g_{s}^{2}}{M_{s}^{2}}+ \frac{g_{t}^{2}}{M_{t}^{2}})\overline{n}n+ (\frac{g_{s}^{2}}{M_{s}^{2}}- \frac{g_{t}^{2}}{M_{t}^{2}})\overline{p}p)
\end{array}
\end{equation}

\noindent which clearly gives Eq (\ref{sisig}) with $f_{p}= g_{s}^{2}/M_{s}^{2}- g_{t}^{2}/M_{t}^{2}$ and $f_{n}= g_{s}^{2}/M_{s}^{2}+g_{t}^{2}/M_{t}^{2}$, and assumes the lightest dark spinor constituting the DM is $\chi$. Of course, if the lightest dark spinor has positive isospin ($\psi$), $f_{p}= g_{s}^{2}/M_{s}^{2}+ g_{t}^{2}/M_{t}^{2}$ and $f_{n}= g_{s}^{2}/M_{s}^{2}- g_{t}^{2}/M_{t}^{2}$.

The terms $g_{t}\vec{\phi} \cdot\overline{\Psi}\vec{\sigma}\Psi$ and $g_{s}\rho\overline{\Psi} \Psi$ in Eq. (\ref{lagra}) provide annihilation channels of the dark fermions $\Psi$ to the $\rho$ and $\vec{\phi}$ scalars, as well as all standard model particles lighter than $\chi$ or $\psi$ to which $\rho$ and $\vec{\phi}$ may couple. Moreover, whenever $M_{t,s}\gg M_{\psi,\chi}$, the effective lagrangian for annihilation to any isospin doublet $\Lambda$ has the same form of Eq. (\ref{efflagra}), i.e. $\left(\frac{g_{t}}{M_{t}}\right)^{2}\overline{\Psi} \vec{\sigma}\Psi\cdot \overline{\Lambda}\vec{\sigma}\Lambda +\left(\frac{g_{s}}{M_{s}}\right)^{2} \overline{\Psi} \Psi\overline{\Lambda}\Lambda$, which has enough free parameters to easily provide the correct DM relic density in a way analogous to what shown in Ref. \cite{wimpless}. E.g., if the lightest dark fermion is $\chi$, the annihilation cross section of $\chi$ to an isospin singlet $\mu$ ($\overline{\chi}\chi\rightarrow \overline{\mu}\mu$) in the center of mass reference frame is:

\begin{equation}
\sigma\propto \frac{g_{s}^{2}M_{\chi}(M_{\chi}-M_{\mu})}{M_{s}^{4}}\approx \frac{g_{s}^{2}M_{\chi}^{2}}{M_{s}^{4}}
\end{equation}

\noindent where the last expression is valid for $M_{\mu}\ll M_{\chi}$. As for the impact of the model-blind SI analysis with $f_{p}\neq f_{n}$ on current direct dark matter results, it has been shown in the literature \cite{cogexe,ivdm} that by simply choosing $f_{p}/f_{n} \approx$-(Xe neutrons)/(Xe protons)$\approx -78/54 \approx -1.4$ it is possible to suppress the zero momentum transfer cross section of the Xe isotopes enough to make the exclusions from XENON compatible with the CoGeNT positive signal, and, at the same time, reconcile the 2 annually modulated positive signals currently available. This, of course, applies to whatever model can yield essentially arbitrary $f_{p,n}$, independently on whether they violate isospin conservation/symmetry or not. It is therefore possible that the DM has an isospin dependent interaction ($f_{p}\neq f_{n}$) with the nucleons without even violating isospin conservation, and these scenarios can better fit the known observations and exclusions than the more theoretically biased $f_{p}= f_{n}$ models. Indeed, at present, the most appropriate approach to DM data analysis and interpretation is still to simply drop the $f_{p}= f_{n}$ restriction, waiting for the more recent positive signals to become solid (CoGeNT admittedly still has to perform many systematic studies, and CRESST \cite{CRESST} doesn't yet have a modulation) to extract from the data clues on which theoretical model best fits the observations.

\begin{figure*}
\begin{center}
\includegraphics[width=.49\linewidth]{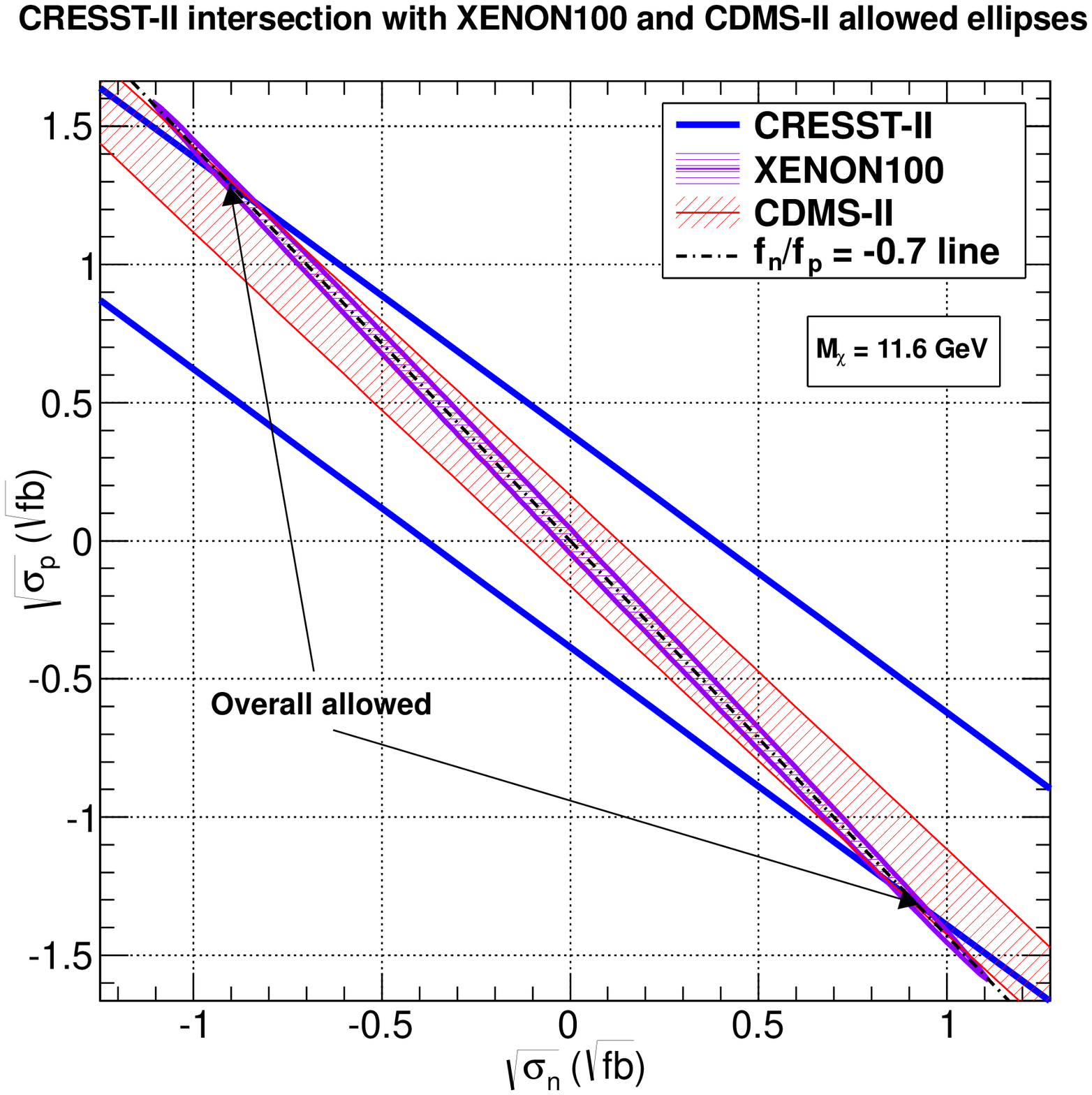}
\includegraphics[width=.49\linewidth]{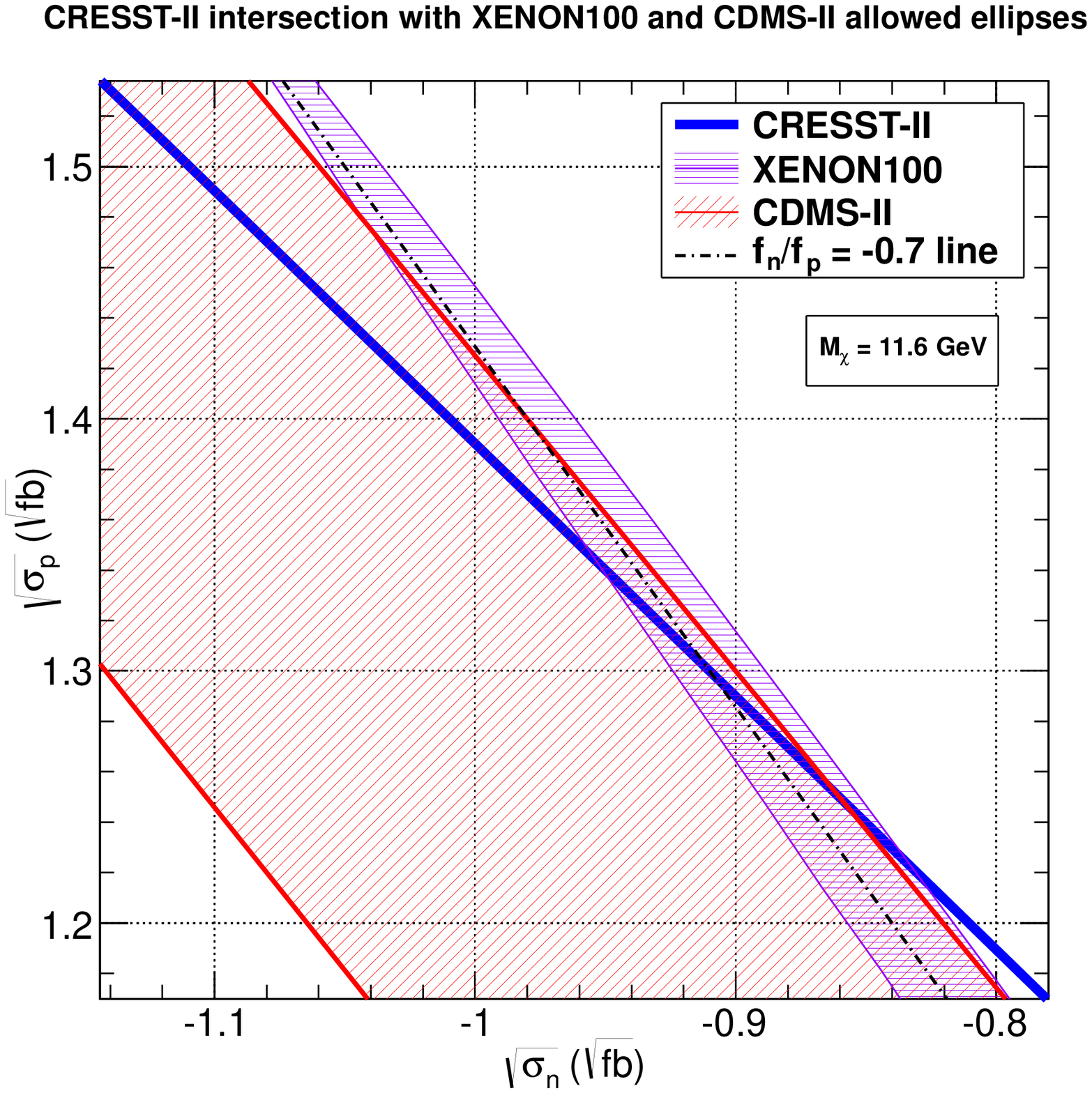}
\caption{CRESST-II (blue thick line), XENON100 (violet horizontally hatched area) and CDMS-II (red hatched area) for M$_{\chi}=11.6$ GeV (left), and detail of the upper left intersection region of the 3 experiments (right). As the overlap region of the red ellipsis allowed by CDMS-II and the violet ellipsis allowed by XENON100 intersects the CRESST-II elliptical contour, there are values of $f_{p}$ and $f_{n}$ for which the 2 results are compatible. Not surprisingly (see text), these values have a ratio $f_{n}/f_{p}$ close to -0.7 (dash-dotted line) which maximally suppresses the SI sensitivity of the Xe isotopes.}
\label{crexecompa}
\end{center}
\end{figure*}

As a new example of how $f_{p}\neq f_{n}$ can reconcile positive signals and exclusions (provided that the detector targets are different), and of how the data from different experiments can tell us which the possible values of $f_{p,n}$ are, let us consider the latest DM signal claimed by CRESST-II, focusing on the M2 candidate of Ref. \cite{CRESST}, which, in the standard $f_{p}= f_{n}$ analysis, has $M_{\chi}=11.6$ GeV and $\sigma=3.7\times 10^{-2}$ fb, in clear contrast with both the $\approx 3.3\times 10^{-4}$ fb limit of the XENON100 and the $\approx 5.2\times 10^{-3}$ fb of the CDMS-II exclusions \cite{xenon100,cdmsII,dmtools}. Now, if an experiment reports an $f_{p}= f_{n}$ cross section $\sigma _{N}$ for the nucleon, it means its isotope (Z,A) had a zero momentum transfer DM cross section  

\begin{equation}
\label{sigmaa}
\sigma_{Z}^{A}=\sigma _{N}A^{2}(\frac{\mu_{A}} {\mu_{p}})^{2}
\end{equation}

\noindent where $\mu_{A,p}$ are the DM-nucleus and DM-proton reduced masses, and the small difference between the 2 nucleon masses has been neglected. But from Eq. (\ref{sisig}) it is clear that

\begin{equation}
\label{convert}
\left\{\begin{array}{l}
\sigma_{p}=\frac{\sigma_{Z}^{A}}{[Z+f_{n}/f_{p}(A-Z)]^{2}}(\frac{\mu_{p}} {\mu_{A}})^{2}=\\
\sigma_{N}\frac{A^{2}}{[Z+f_{n}/f_{p}(A-Z)]^{2}}\\
\sigma_{n}=\sigma_{N}\frac{A^{2}}{f_{p}/f_{n}Z+(A-Z)^{2}}
\end{array}\right .
\end{equation}

\noindent where, of course, $\sigma_{n}$ is the neutron, and $\sigma_{N}$ the nucleon cross section.
Picking the most abundant isotopes of each elements in the respective detectors, and using the above values of $\sigma _{N}$, it is easy to obtain from Eq. (\ref{convert}) approximate proton and neutron cross sections for XENON100 and CRESST-II in the scenario $f_{n}/f_{p}=-0.7$ of Ref.s \cite{cogexe,ivdm}: $\sigma_{p} \approx 16$ fb for $^{132}$Xe of XENON100, $\sigma_{p}\approx 4.2$ fb for  $^{74}$Ge of CDMS-II, $\sigma_{p}\approx 1.6$ fb for both $^{16}$O and $^{40}$Ca of CRESST-II. Clearly, in this Xe sensitivity suppressing scenario and this rough approximation, the signal of CRESST-II is not excluded by XENON100 nor CDMS-II. In fact, also the neutron interaction is not excluded:
$\sigma_{n} \approx 8$ fb for Xe$^{132}$ of XENON100, $\sigma_{n}\approx 0.8$ fb for both O$^{16}$ and Ca$^{40}$ of CRESST-II, and the $\sigma_{n}\approx 2$ fb for  $^{74}$Ge of CDMS-II is still compatible with the 11.6 GeV candidate of CRESST-II, though the maximal suppression for $^{74}$Ge occurs at $f_{n}/f_{p}=-0.76$, equal and opposite to the $Z/(A-Z)$ ratio of $^{74}$Ge. This is why, in this rough estimate and specific scenario, the CDMS-II limit seems closer to the CRESST measurement than XENON100's.
The reason for the identity of $\sigma_{p,n}$ of $^{16}$O and $^{40}$Ca in CRESST-II is that, for all isotopes with $A=2Z$, $A^{2}/[f_{p}/f_{n}Z+(A-Z)]^{2}=A^{2}/[Z+f_{n}/f_{p}(A-Z)]^{2}=4/(1+f_{n}/f_{p})^{2}$.
 
To take a more rigorous approach, I'll follow Ref. \cite{me}, with some modifications. The ellipses displayed in Fig. \ref{crexecompa} are given by Eq.s (6,7) of Ref. \cite{me}, with the following differences:

\begin{itemize}
  \item[--] instead of the coupling strengths, the adopted variables are the signed square roots of $\sigma_{p,n}$, so that the sign ambiguity in the first of Eq.s (7) (Ref. \cite{me}) can be removed, while the right hand term remains 1.
  \item[--] For CRESST-II, the equation is an equality, while for XENON100 and CDMS-II it remains an inequality (limit). This means that XENON100 and CDMS-II allow the interior of the respective ellipses, while CRESST-II affirms the proton and neutron cross sections for $M_{\chi}=11.6$ GeV are, within errors, somewhere along its ellipsis.
\end{itemize}

The conversion from $\sigma_N$ to $\sigma_{p,n}$ starts from the standard rate equation:

\begin{equation}
\label{rate}
\frac{dR}{dE}=\frac{\rho}{M_{\chi}}\sum_{A}\frac{\sigma_{A}}{2\mu_{A}^2}FF_{A}^{2}f_{A}I_{A}\approx\frac{\rho}{2M_{\chi}}I\sum_{A}\frac{\sigma_{A}}{\mu_{A}^2}f_{A}
\end{equation}

\noindent where $dR/dE$ is the rate per unit target mass, $\sigma_{A}$ the scattering cross section of isotope A, $\mu_{A}$ the DM particle--nucleus reduced mass, $FF_{A}^{2}$ the nuclear form factor, $f_{A}$ the abundance of isotope A in the target's composition and $I_{A}=\int f(v)/vdv$ the quasi moment -1 of the DM velocity distribution relative to the detector, which depends on A via the lower integration extremum (minimum incident velocity able to produce the recoil energy E). The rightmost side of Eq. (\ref{rate}) employs the simplifying assumptions of neglecting the weak dependence of $I_{A}$ on A ($I_{A}\approx I$) and that for light DM the momentum transfer is small (so that $FF_{A}^{2} \approx 1$). From Eq. (\ref{sisig}) we can deduce:

\begin{equation}
\label{cesemo}
\frac{\sigma_{A}}{\mu_{A}^2}=\frac{[Z\sqrt{\sigma_{p}}+(A-Z)\sqrt{\sigma_{n}}]^{2}}{\mu_{p}^2} =A^{2}\frac{\sigma_N}{\mu_{p}^2}
\end{equation}

\noindent again containing the approximation $\mu_p \approx \mu_n$, the rightmost side being the value of the center one for the signed $\sqrt{\sigma_{p}}=\sqrt{\sigma_{n}}$, i.e. for $f_p=f_n$. Substituting Eq. (\ref{cesemo}) in the rightmost side of Eq. (\ref{rate}) and dividing by $\rho/2M_{\chi}I\sum_{A}f_{A}A^{2}\frac{\sigma_N}{\mu_{p}^2}$, we obtain:

\begin{equation}
\sum_{A}\frac{f_{A}}{\sum_{A}f_{A}A^{2}}\frac{[\sqrt{\sigma_p}Z+ \sqrt{\sigma_n}(A-Z)]^{2}}{\sigma_{N}}=1
\end{equation}

\noindent which produces the ellipses of Fig. \ref{crexecompa}. The consistency of the above approximations has been verified by intersecting the ellipses with the $\sqrt{\sigma_p}= \sqrt{\sigma_n}$ line, finding that the coordinates of the intersections are indeed the square roots of the $\sigma_N$ reported by the experiments for $M_{\chi}=11.6$ GeV.

\begin{figure}
\begin{center}
\includegraphics[width=\linewidth]{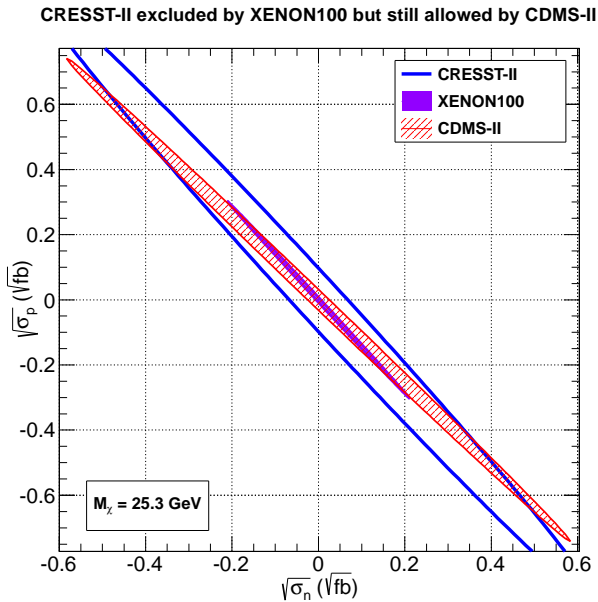}
\caption{CRESST-II (blue thick line), XENON100 (violet solid area) and CDMS-II (red hatched area)  for $M_{\chi}=25.3$ GeV. As XENON100's ellipsis is entirely contained in CRESST-II's, the M1 candidate is incompatible with XENON100's exclusions at this DM mass for any SI couplings.  CDMS-II is instead still compatible with CRESST-II for $f_{p}/f_{n} \sim -1.2$.}
\label{crexecompaM1}
\end{center}
\end{figure}

XENON100 and CDMS-II allow, for $M_{\chi}=11.6$ GeV, the inside of the respective ellipses, while CDMS-II claims the most likely values of $\sqrt{\sigma_p}$ and $\sqrt{\sigma_n}$ are along its elliptical contour. Since the latter intersects the overlap area of the XENON100 and CDMS-II allowed regions, there are some values of $\sqrt{\sigma_p}$ and $\sqrt{\sigma_n}$ from CRESST-II still allowed by both XENON100 and CDMS-II. The left Fig. \ref{crexecompa} shows the two symmetric areas compatible with the CRESST-II M2 candidate, while the righthand plot shows one of them in detail. Given that the $f_{n}/f_{p} = -0.7$ crosses the middle of the two symmetric areas still allowed, the $f_{n}/f_{p}$ ratio is constrained to values near $\sim-0.7$.

For completeness, let us consider also CRESST-II's candidate M1 ($M_{\chi}=25.3$ GeV, 
$\approx1.6\times 10^{-3}$ fb, vs the $\approx1.2 \times 10^{-5}$ fb limit of XENON100 and $\approx 2\times 10^{-4}$ fb of CDMS-II in the $f_{p}=f_{n}$ scenario), which has a 69\% contribution from tungsten. From the latter circumstance, it is reasonable to expect that XENON100 constrains CRESST-II more strongly than at 11.6 GeV, since the $(A-Z)/Z$ ratios of tungsten's isotopes are closer to those of xenon than they are for oxygen and calcium, so that scenarios suppressing XENON100's sensitivity to the M1 candidate would largely suppress also CRESST-II's sensitivity. Fig. \ref{crexecompaM1} confirms this expectation: XENON100's exclusions are incompatible, at maximum likelyhood, with the M1 candidate, for every choice of $f_{p}/f_{n}$. CDMS-II, instead, is still compatible, for $f_{p}/f_{n} \sim -1.2$, with M1. Exploring the full $2\sigma$ M1 region of Ref. \cite{CRESST} is beyond the scope of this paper, and would probably require the detailed knowledge of CRESST-II, XENON100 and CDMS-II data that only the respective collaborations have.

In conclusion, it is still easy to devise simple models of DM where the DM particle couples with the two nucleons with different coupling strengths, even without violating isospin conservation, and can yield the correct cosmological relic abundance of DM. This circumstance makes it necessary, for a correct comparison of experimental results some of which are positive signals, to analyze (or reanalyze) the data without assuming $f_{p}=f_{n}$. In fact, while for comparing the achieved sensitivities of experiments without positive signal it makes sense to pick the scenario in which all experiments are most sensitive, positive signals may well arise from a scenario where this is not the case. The latest example of the above is the apparent contrast between the recent CRESST-II event excess and current exclusions, like those of XENON100 and CDMS-II, which are currently compared in a model biased fashion. Dropping the $f_{p}=f_{n}$ bias shows that only the M1 SI candidate is excluded for all $f_{p,n}$, while for M2 $f_{p}/f_{n}$ is constrained to values close to -1.4.

\end{document}